\documentclass[draftclsnofoot,onecolumn]{IEEEtran} 	
\usepackage{amsmath,amssymb}
\usepackage{amsthm}
\usepackage{color}
\usepackage[mathscr]{eucal}
\usepackage{graphics,graphicx,multicol}
\usepackage{epsfig}
\usepackage{enumerate}
\usepackage{subfigure}
\usepackage{algorithmic}
\usepackage[section]{algorithm}
\usepackage{morefloats}
\usepackage{enumerate}
\usepackage{subfigure}
\usepackage{multirow}
\usepackage{xfrac}
\usepackage{array}
\usepackage{pstricks, pst-node, pst-plot, pst-circ}
\usepackage{pst-3dplot,pstricks-add,pst-solides3d,pst-grad}
\usepackage{moredefs}
\usepackage{cite}
\usepackage{courier}

\begin{document}

\title{Context-Aware Resource Allocation in Cellular Networks}
\author{Ahmed Abdelhadi and Charles Clancy \\
Hume Center, Virginia Tech\\
\{aabdelhadi, tcc\}@vt.edu
}
\maketitle

\begin{abstract}

We define and propose a resource allocation architecture for cellular networks. The architecture combines content-aware, time-aware and location-aware resource allocation for next generation broadband wireless systems. The architecture ensures content-aware resource allocation by prioritizing real-time applications users over delay-tolerant applications users when allocating resources. It enables time-aware resource allocation via traffic-dependent pricing that varies during different hours of day (e.g. peak and off-peak traffic hours). Additionally, location-aware resource allocation is integrable in this architecture by including carrier aggregation of various frequency bands. The context-aware resource allocation is an optimal and flexible architecture that can be easily implemented in practical cellular networks. We highlight the advantages of the proposed network architecture with a discussion on the future research directions for context-aware resource allocation architecture. We also provide 
experimental results to illustrate a general proof of concept for this new architecture.

\end{abstract}


\providelength{\AxesLineWidth}       \setlength{\AxesLineWidth}{0.5pt}%
\providelength{\plotwidth}           \setlength{\plotwidth}{10cm}
\providelength{\LineWidth}           \setlength{\LineWidth}{0.7pt}%
\providelength{\LineWidthTwo}        \setlength{\LineWidthTwo}{1.7pt}%
\providelength{\MarkerSize}          \setlength{\MarkerSize}{3.5pt}%
\newrgbcolor{GridColor}{0.8 0.8 0.8}%
\newrgbcolor{GridColor2}{0.5 0.5 0.5}%

\section{Introduction}\label{sec:intro}

Wireless broadband systems are witnessing rapid growth in both the number of subscribers and the traffic volume per subscriber. More smartphone users are using web-based services such as video streaming, e.g. YouTube, and social networking, e.g. Facebook. These applications require broadband access due to multimedia-rich content. Given the limited available cellular spectrum and the aforementioned increasing demand, the need for an efficient resource allocation algorithm is of paramount importance for improving the quality of service (QoS).

The user applications running on smartphones can be divided into real-time applications, e.g. voice-over-IP (VoIP), video streaming, etc., and delay tolerant applications, e.g. application updates, emails, etc. The real-time applications are given allocation priority over the delay-tolerant applications due to latency constraints. Meanwhile, users running delay-tolerant application shouldn't be deprived from resources (i.e. no user in the network is dropped). Hence, we aim for an efficient content-aware resource allocation under the limited cellular spectrum that ensures priority to real-time applications without dropping users with delay-tolerant applications.

In addition, the cellular network traffic during peak-traffic hours can be ten times more that off-peak traffic hours \cite{pricing_survey}. This causes congestion problems that result in (a) lower QoS for subscribers and therefore wireless service providers (WSP)s have to (b) deploy more equipments and base stations to meet the increasing demand and therefore the cost of network operation increases leading to (c) higher pricing by WSP to users to cover the cost of operation. Therefore, there is a real need for a time-aware pricing model that overcome the congestion during peak traffic hours to solve the aforementioned issues. 

This article aims to illustrate that context-aware resource allocation architecture with frequency reuse and its benefits to future cellular networks. The architecture considers users running real-time applications and users running delay-tolerant applications. The users running different content experience different resource allocation, due to the proposed content-aware resource allocation policy. The resource allocation optimization problem is formulated to ensure fair utility percentage allocated for active users with the available evolved-NodeB (eNodeB) spectrum resources. Therefore, the context-aware resource allocation algorithm gives priority to real-time application users over delay-tolerant application users. In addition, the optimization problem formulation guarantees that all users are assigned a fraction of the available spectrum, as the eNodeB should provide a minimum QoS for all the users subscribing for the mobile service. This allocation policy intrinsically provides time-aware pricing that 
charges mobile users based on their usage time of day (i.e. peak and off-peak traffic hours).

\subsection{Related Work}\label{sec:related}
The area of resource allocation optimization has received significant interest since the seminal network utility maximization problem presented in \cite{kelly98ratecontrol}. The network utility maximization problem allocates the resources among users optimally based on bandwidth proportional fairness by using Lagrange multiplier methods of optimization theory. An iterative algorithm based on the dual problem has been proposed to solve the resource allocation optimization problem in \cite{Low99optimizationflow}. The applications considered in early research work, as in \cite{kelly98ratecontrol} and \cite{Low99optimizationflow}, are only delay-tolerant Internet traffic for wired communication networks. However, for current cellular networks both real-time and delay-tolerant applications are considered. The content-aware resource allocation architecture optimally allocates resources for these heterogeneous applications. 

For earlier studies on the single carrier content-aware resource allocation, we refer to \cite{Ahmed_Utility1}. Time-aware resource pricing was introduced in \cite{Ahmed_Utility2} for single cell model. The location-aware resource allocation with carrier aggregation is studied in \cite{Ahmed_Utility4, Haya_Utility1}. The context-aware resource allocation with prioritization of mobile users based on their subscription is investigated in \cite{Ahmed_Utility3}. The context-aware resource allocation with guaranteed bit rate (GBR) to mobile users running specific services is discussed in \cite{Haya_Utility2}. 

The article is organized as follows. In the next section, the context-aware resource allocation network architecture is presented along with the allocation policy. The following section provides the distribution of the context-aware resource allocation problem into cellular network entities subproblems. After that, the distributed optimal context-aware resource allocation algorithm running on various cellular network entities is presented. Then, the algorithm is investigated using simulations and the possible future extensions to the current architecture are discussed. The final section concludes the article.

\section{Context-Aware Resource Allocation Network Architecture}\label{sec:Problem_formulation}

In this section, we define the problem under investigation. We break it into utility functions, system model, resource allocation policy, and resource allocation optimization problem.

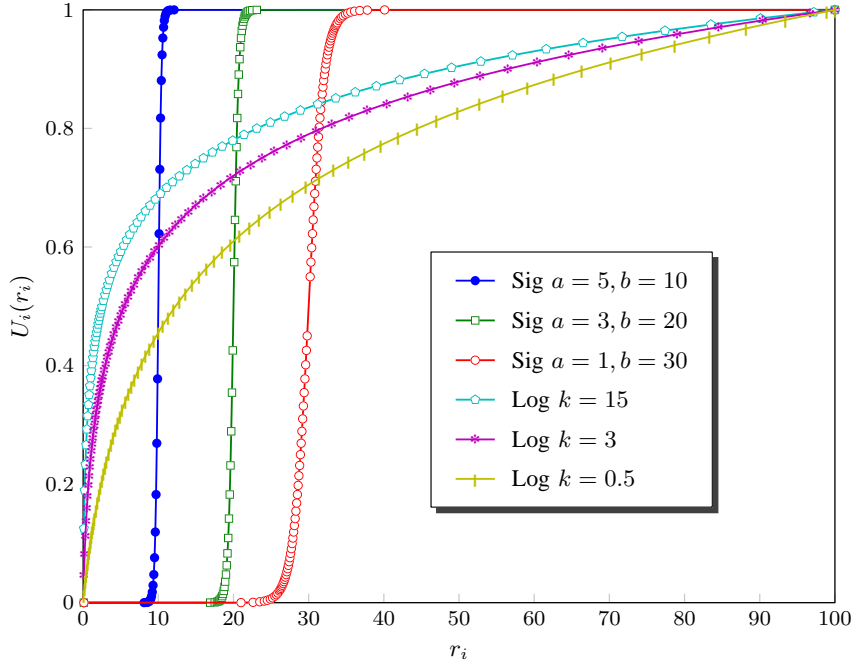
\begin{figure}[]
\centering
%
\psset{xunit=0.010000\plotwidth,yunit=0.788710\plotwidth}%
\begin{pspicture}(-11.059908,-0.111111)(102.764977,1.023392)%
\psline[linewidth=\AxesLineWidth,linecolor=GridColor](0.000000,0.000000)(0.000000,0.015215)
\psline[linewidth=\AxesLineWidth,linecolor=GridColor](10.000000,0.000000)(10.000000,0.015215)
\psline[linewidth=\AxesLineWidth,linecolor=GridColor](20.000000,0.000000)(20.000000,0.015215)
\psline[linewidth=\AxesLineWidth,linecolor=GridColor](30.000000,0.000000)(30.000000,0.015215)
\psline[linewidth=\AxesLineWidth,linecolor=GridColor](40.000000,0.000000)(40.000000,0.015215)
\psline[linewidth=\AxesLineWidth,linecolor=GridColor](50.000000,0.000000)(50.000000,0.015215)
\psline[linewidth=\AxesLineWidth,linecolor=GridColor](60.000000,0.000000)(60.000000,0.015215)
\psline[linewidth=\AxesLineWidth,linecolor=GridColor](70.000000,0.000000)(70.000000,0.015215)
\psline[linewidth=\AxesLineWidth,linecolor=GridColor](80.000000,0.000000)(80.000000,0.015215)
\psline[linewidth=\AxesLineWidth,linecolor=GridColor](90.000000,0.000000)(90.000000,0.015215)
\psline[linewidth=\AxesLineWidth,linecolor=GridColor](100.000000,0.000000)(100.000000,0.015215)
\psline[linewidth=\AxesLineWidth,linecolor=GridColor](0.000000,0.000000)(1.200000,0.000000)
\psline[linewidth=\AxesLineWidth,linecolor=GridColor](0.000000,0.200000)(1.200000,0.200000)
\psline[linewidth=\AxesLineWidth,linecolor=GridColor](0.000000,0.400000)(1.200000,0.400000)
\psline[linewidth=\AxesLineWidth,linecolor=GridColor](0.000000,0.600000)(1.200000,0.600000)
\psline[linewidth=\AxesLineWidth,linecolor=GridColor](0.000000,0.800000)(1.200000,0.800000)
\psline[linewidth=\AxesLineWidth,linecolor=GridColor](0.000000,1.000000)(1.200000,1.000000)
{ \footnotesize 
\rput[t](0.000000,-0.015215){$0$}
\rput[t](10.000000,-0.015215){$10$}
\rput[t](20.000000,-0.015215){$20$}
\rput[t](30.000000,-0.015215){$30$}
\rput[t](40.000000,-0.015215){$40$}
\rput[t](50.000000,-0.015215){$50$}
\rput[t](60.000000,-0.015215){$60$}
\rput[t](70.000000,-0.015215){$70$}
\rput[t](80.000000,-0.015215){$80$}
\rput[t](90.000000,-0.015215){$90$}
\rput[t](100.000000,-0.015215){$100$}
\rput[r](-1.200000,0.000000){$0$}
\rput[r](-1.200000,0.200000){$0.2$}
\rput[r](-1.200000,0.400000){$0.4$}
\rput[r](-1.200000,0.600000){$0.6$}
\rput[r](-1.200000,0.800000){$0.8$}
\rput[r](-1.200000,1.000000){$1$}
} 
\psframe[linewidth=\AxesLineWidth,dimen=middle](0.000000,0.000000)(100.000000,1.000000)
{ \small 
\rput[b](50.000000,-0.111111){
\begin{tabular}{c}
$r_{i}$\\
\end{tabular}
}
\rput[t]{90}(-11.059908,0.500000){
\begin{tabular}{c}
$U_i(r_{i})$\\
\end{tabular}
}
} 
\newrgbcolor{color18.0052}{0  0  1}
\psline[plotstyle=line,linejoin=1,showpoints=false,dotstyle=*,dotsize=\MarkerSize,linestyle=solid,linewidth=\LineWidth,linecolor=color18.0052]
(100.000000,1.000000)(100.000000,1.000000)
\psline[plotstyle=line,linejoin=1,showpoints=true,dotstyle=*,dotsize=\MarkerSize,linestyle=solid,linewidth=\LineWidth,linecolor=color18.0052](0.100000,0.000000)(8.100000,0.000075)(8.400000,0.000335)(8.600000,0.000911)(8.700000,0.001501)
(8.800000,0.002473)(8.900000,0.004070)(9.000000,0.006693)(9.100000,0.010987)(9.200000,0.017986)
(9.300000,0.029312)(9.400000,0.047426)(9.500000,0.075858)(9.600000,0.119203)(9.700000,0.182426)
(9.800000,0.268941)(9.900000,0.377541)(10.100000,0.622459)(10.200000,0.731059)(10.300000,0.817574)
(10.400000,0.880797)(10.500000,0.924142)(10.600000,0.952574)(10.700000,0.970688)(10.800000,0.982014)
(10.900000,0.989013)(11.000000,0.993307)(11.100000,0.995930)(11.200000,0.997527)(11.300000,0.998499)
(11.400000,0.999089)(11.600000,0.999665)(12.100000,0.999972)(100.000000,1.000000)
\newrgbcolor{color19.0048}{0         0.5           0}
\psline[plotstyle=line,linejoin=1,showpoints=false,dotstyle=Bsquare,dotsize=\MarkerSize,linestyle=solid,linewidth=\LineWidth,linecolor=color19.0048]
(100.000000,1.000000)(100.000000,1.000000)
\psline[plotstyle=line,linejoin=1,showpoints=true,dotstyle=Bsquare,dotsize=\MarkerSize,linestyle=solid,linewidth=\LineWidth,linecolor=color19.0048]
(0.100000,0.000000)(16.900000,0.000091)(17.500000,0.000553)(17.800000,0.001359)(18.000000,0.002473)
(18.100000,0.003335)(18.200000,0.004496)(18.300000,0.006060)(18.400000,0.008163)(18.500000,0.010987)
(18.600000,0.014774)(18.700000,0.019840)(18.800000,0.026597)(18.900000,0.035571)(19.000000,0.047426)
(19.100000,0.062973)(19.200000,0.083173)(19.300000,0.109097)(19.400000,0.141851)(19.500000,0.182426)
(19.600000,0.231475)(19.700000,0.289050)(19.800000,0.354344)(19.900000,0.425557)(20.100000,0.574443)
(20.200000,0.645656)(20.300000,0.710950)(20.400000,0.768525)(20.500000,0.817574)(20.600000,0.858149)
(20.700000,0.890903)(20.800000,0.916827)(20.900000,0.937027)(21.000000,0.952574)(21.100000,0.964429)
(21.200000,0.973403)(21.300000,0.980160)(21.400000,0.985226)(21.500000,0.989013)(21.600000,0.991837)
(21.700000,0.993940)(21.800000,0.995504)(21.900000,0.996665)(22.000000,0.997527)(22.200000,0.998641)
(22.500000,0.999447)(23.100000,0.999909)(100.000000,1.000000)
\newrgbcolor{color20.0048}{1  0  0}
\psline[plotstyle=line,linejoin=1,showpoints=false,dotstyle=Bo,dotsize=\MarkerSize,linestyle=solid,linewidth=\LineWidth,linecolor=color20.0048]
(100.000000,1.000000)(100.000000,1.000000)
\psline[plotstyle=line,linejoin=1,showpoints=true,dotstyle=Bo,dotsize=\MarkerSize,linestyle=solid,linewidth=\LineWidth,linecolor=color20.0048]
(0.100000,0.000000)(21.000000,0.000123)(22.600000,0.000611)(23.500000,0.001501)(24.100000,0.002732)
(24.500000,0.004070)(24.900000,0.006060)(25.200000,0.008163)(25.500000,0.010987)(25.700000,0.013387)
(25.900000,0.016302)(26.100000,0.019840)(26.200000,0.021881)(26.300000,0.024127)(26.400000,0.026597)
(26.500000,0.029312)(26.600000,0.032295)(26.700000,0.035571)(26.800000,0.039166)(26.900000,0.043107)
(27.000000,0.047426)(27.100000,0.052154)(27.200000,0.057324)(27.300000,0.062973)(27.400000,0.069138)
(27.500000,0.075858)(27.600000,0.083173)(27.700000,0.091123)(27.800000,0.099750)(27.900000,0.109097)
(28.000000,0.119203)(28.100000,0.130108)(28.200000,0.141851)(28.300000,0.154465)(28.400000,0.167982)
(28.500000,0.182426)(28.600000,0.197816)(28.700000,0.214165)(28.800000,0.231475)(28.900000,0.249740)
(29.000000,0.268941)(29.100000,0.289050)(29.200000,0.310026)(29.300000,0.331812)(29.400000,0.354344)
(29.500000,0.377541)(29.600000,0.401312)(29.700000,0.425557)(29.800000,0.450166)(30.200000,0.549834)
(30.300000,0.574443)(30.400000,0.598688)(30.500000,0.622459)(30.600000,0.645656)(30.700000,0.668188)
(30.800000,0.689974)(30.900000,0.710950)(31.000000,0.731059)(31.100000,0.750260)(31.200000,0.768525)
(31.300000,0.785835)(31.400000,0.802184)(31.500000,0.817574)(31.600000,0.832018)(31.700000,0.845535)
(31.800000,0.858149)(31.900000,0.869892)(32.000000,0.880797)(32.100000,0.890903)(32.200000,0.900250)
(32.300000,0.908877)(32.400000,0.916827)(32.500000,0.924142)(32.600000,0.930862)(32.700000,0.937027)
(32.800000,0.942676)(32.900000,0.947846)(33.000000,0.952574)(33.100000,0.956893)(33.200000,0.960834)
(33.300000,0.964429)(33.400000,0.967705)(33.500000,0.970688)(33.600000,0.973403)(33.700000,0.975873)
(33.800000,0.978119)(34.000000,0.982014)(34.200000,0.985226)(34.400000,0.987872)(34.600000,0.990048)
(34.900000,0.992608)(35.200000,0.994514)(35.600000,0.996316)(36.100000,0.997762)(36.800000,0.998887)
(37.800000,0.999590)(40.100000,0.999959)(100.000000,1.000000)
\newrgbcolor{color21.0048}{0        0.75        0.75}
\psline[plotstyle=line,linejoin=1,showpoints=false,dotstyle=Bpentagon,dotsize=\MarkerSize,linestyle=solid,linewidth=\LineWidth,linecolor=color21.0048]
(100.000000,1.000000)(100.000000,1.000000)
\psline[plotstyle=line,linejoin=1,showpoints=true,dotstyle=Bpentagon,dotsize=\MarkerSize,linestyle=solid,linewidth=\LineWidth,linecolor=color21.0048]
(0.100000,0.125281)(0.200000,0.189543)(0.300000,0.233084)(0.400000,0.266057)(0.500000,0.292603)
(0.600000,0.314824)(0.700000,0.333933)(0.800000,0.350696)(0.900000,0.365626)(1.000000,0.379086)
(1.100000,0.391338)(1.200000,0.402582)(1.300000,0.412971)(1.400000,0.422627)(1.500000,0.431645)
(1.600000,0.440105)(1.700000,0.448072)(1.800000,0.455600)(1.900000,0.462735)(2.000000,0.469516)
(2.100000,0.475977)(2.200000,0.482146)(2.300000,0.488049)(2.400000,0.493707)(2.500000,0.499141)
(2.700000,0.509400)(2.900000,0.518943)(3.100000,0.527863)(3.300000,0.536237)(3.500000,0.544127)
(3.800000,0.555169)(4.100000,0.565386)(4.400000,0.574892)(4.700000,0.583780)(5.000000,0.592125)
(5.400000,0.602514)(5.800000,0.612169)(6.200000,0.621187)(6.700000,0.631683)(7.200000,0.641430)
(7.700000,0.650528)(8.300000,0.660703)(8.900000,0.670172)(9.600000,0.680450)(10.300000,0.690009)
(11.100000,0.700173)(11.900000,0.709633)(12.800000,0.719548)(13.800000,0.729781)(14.900000,0.740219)
(16.000000,0.749915)(17.200000,0.759764)(18.500000,0.769689)(19.900000,0.779628)(21.400000,0.789533)
(23.100000,0.799953)(24.900000,0.810184)(26.800000,0.820212)(28.900000,0.830502)(31.200000,0.840949)
(33.600000,0.851060)(36.200000,0.861231)(39.000000,0.871400)(42.100000,0.881840)(45.400000,0.892143)
(49.000000,0.902561)(52.900000,0.913018)(57.100000,0.923452)(61.600000,0.933812)(66.500000,0.944266)
(71.700000,0.954550)(77.400000,0.965000)(83.500000,0.975363)(90.100000,0.985756)(97.200000,0.996120)
(100.000000,1.000000)
\newrgbcolor{color22.0046}{0.75           0        0.75}
\psline[plotstyle=line,linejoin=1,showpoints=false,dotstyle=Basterisk,dotsize=\MarkerSize,linestyle=solid,linewidth=\LineWidth,linecolor=color22.0046]
(100.000000,1.000000)(100.000000,1.000000)
\psline[plotstyle=line,linejoin=1,showpoints=true,dotstyle=Basterisk,dotsize=\MarkerSize,linestyle=solid,linewidth=\LineWidth,linecolor=color22.0046]
(0.100000,0.045971)(0.200000,0.082354)(0.300000,0.112466)(0.400000,0.138154)(0.500000,0.160552)
(0.600000,0.180410)(0.700000,0.198244)(0.800000,0.214430)(0.900000,0.229246)(1.000000,0.242907)
(1.100000,0.255579)(1.200000,0.267396)(1.300000,0.278466)(1.400000,0.288878)(1.500000,0.298706)
(1.600000,0.308012)(1.700000,0.316848)(1.800000,0.325261)(1.900000,0.333287)(2.000000,0.340962)
(2.100000,0.348315)(2.200000,0.355372)(2.300000,0.362156)(2.400000,0.368686)(2.500000,0.374982)
(2.600000,0.381060)(2.800000,0.392617)(3.000000,0.403459)(3.200000,0.413669)(3.400000,0.423316)
(3.600000,0.432460)(3.800000,0.441151)(4.100000,0.453428)(4.400000,0.464901)(4.700000,0.475669)
(5.000000,0.485813)(5.300000,0.495402)(5.600000,0.504493)(6.000000,0.515925)(6.400000,0.526656)
(6.800000,0.536767)(7.300000,0.548638)(7.800000,0.559755)(8.300000,0.570209)(8.900000,0.581981)
(9.500000,0.593013)(10.100000,0.603391)(10.800000,0.614769)(11.500000,0.625454)(12.300000,0.636916)
(13.100000,0.647675)(14.000000,0.659038)(15.000000,0.670855)(16.000000,0.681925)(17.100000,0.693345)
(18.300000,0.705009)(19.600000,0.716826)(20.900000,0.727896)(22.300000,0.739084)(23.800000,0.750328)
(25.400000,0.761576)(27.100000,0.772785)(29.000000,0.784519)(31.000000,0.796076)(33.100000,0.807443)
(35.400000,0.819100)(37.800000,0.830491)(40.400000,0.842048)(43.200000,0.853696)(46.200000,0.865373)
(49.400000,0.877027)(52.800000,0.888614)(56.500000,0.900409)(60.400000,0.912039)(64.600000,0.923755)
(69.100000,0.935496)(73.900000,0.947209)(79.000000,0.958851)(84.500000,0.970596)(90.400000,0.982378)
(96.700000,0.994140)(100.000000,1.000000)
\newrgbcolor{color23.0046}{0.75        0.75           0}
\psline[plotstyle=line,linejoin=1,showpoints=false,dotstyle=B|,dotsize=\MarkerSize,linestyle=solid,linewidth=\LineWidth,linecolor=color23.0046]
(100.000000,1.000000)(100.000000,1.000000)
\psline[plotstyle=line,linejoin=1,showpoints=true,dotstyle=B|,dotsize=\MarkerSize,linestyle=solid,linewidth=\LineWidth,linecolor=color23.0046]
(0.100000,0.012409)(0.200000,0.024241)(0.300000,0.035546)(0.400000,0.046371)(0.500000,0.056753)
(0.600000,0.066728)(0.700000,0.076327)(0.800000,0.085577)(0.900000,0.094502)(1.000000,0.103124)
(1.100000,0.111463)(1.200000,0.119538)(1.300000,0.127365)(1.400000,0.134957)(1.500000,0.142330)
(1.700000,0.156463)(1.900000,0.169852)(2.100000,0.182572)(2.300000,0.194685)(2.500000,0.206248)
(2.700000,0.217308)(2.900000,0.227906)(3.100000,0.238081)(3.400000,0.252618)(3.700000,0.266370)
(4.000000,0.279415)(4.300000,0.291824)(4.600000,0.303656)(4.900000,0.314962)(5.200000,0.325786)
(5.600000,0.339537)(6.000000,0.352583)(6.400000,0.364992)(6.800000,0.376824)(7.300000,0.390879)
(7.800000,0.404198)(8.300000,0.416854)(8.800000,0.428910)(9.400000,0.442661)(10.000000,0.455707)
(10.600000,0.468116)(11.300000,0.481867)(12.000000,0.494913)(12.800000,0.509046)(13.600000,0.522435)
(14.500000,0.536701)(15.400000,0.550208)(16.400000,0.564421)(17.400000,0.577881)(18.500000,0.591908)
(19.600000,0.605201)(20.800000,0.618953)(22.100000,0.633056)(23.400000,0.646418)(24.800000,0.660064)
(26.300000,0.673915)(27.900000,0.687902)(29.600000,0.701967)(31.400000,0.716056)(33.300000,0.730128)
(35.300000,0.744144)(37.400000,0.758075)(39.600000,0.771894)(41.900000,0.785581)(44.400000,0.799667)
(47.000000,0.813534)(49.800000,0.827667)(52.700000,0.841522)(55.800000,0.855542)(59.100000,0.869663)
(62.600000,0.883830)(66.300000,0.897996)(70.200000,0.912119)(74.300000,0.926167)(78.700000,0.940426)
(83.300000,0.954525)(88.200000,0.968731)(93.400000,0.982986)(98.900000,0.997242)(100.000000,1.000000)
{ \small 
\rput(65.349462,0.371726){%
\psshadowbox[framesep=0pt,linewidth=\AxesLineWidth]{\psframebox*{\begin{tabular}{l}
\Rnode{a1}{\hspace*{0.0ex}} \hspace*{0.3cm} \Rnode{a2}{~~Sig $a=5, b=10$} \\
\Rnode{a3}{\hspace*{0.0ex}} \hspace*{0.3cm} \Rnode{a4}{~~Sig $a=3, b=20$} \\
\Rnode{a5}{\hspace*{0.0ex}} \hspace*{0.3cm} \Rnode{a6}{~~Sig $a=1, b=30$} \\
\Rnode{a7}{\hspace*{0.0ex}} \hspace*{0.3cm} \Rnode{a8}{~~Log $k=15$} \\
\Rnode{a9}{\hspace*{0.0ex}} \hspace*{0.3cm} \Rnode{a10}{~~Log $k=3$} \\
\Rnode{a11}{\hspace*{0.0ex}} \hspace*{0.3cm} \Rnode{a12}{~~Log $k=0.5$} \\
\end{tabular}}
\ncline[linestyle=solid,linewidth=\LineWidth,linecolor=color18.0052]{a1}{a2}
\ncput{\psdot[dotstyle=*,dotsize=\MarkerSize,linecolor=color18.0052]}
\ncline[linestyle=solid,linewidth=\LineWidth,linecolor=color19.0048]{a3}{a4}
\ncput{\psdot[dotstyle=Bsquare,dotsize=\MarkerSize,linecolor=color19.0048]}
\ncline[linestyle=solid,linewidth=\LineWidth,linecolor=color20.0048]{a5}{a6}
\ncput{\psdot[dotstyle=Bo,dotsize=\MarkerSize,linecolor=color20.0048]}
\ncline[linestyle=solid,linewidth=\LineWidth,linecolor=color21.0048]{a7}{a8}
\ncput{\psdot[dotstyle=Bpentagon,dotsize=\MarkerSize,linecolor=color21.0048]}
\ncline[linestyle=solid,linewidth=\LineWidth,linecolor=color22.0046]{a9}{a10}
\ncput{\psdot[dotstyle=Basterisk,dotsize=\MarkerSize,linecolor=color22.0046]}
\ncline[linestyle=solid,linewidth=\LineWidth,linecolor=color23.0046]{a11}{a12}
\ncput{\psdot[dotstyle=B|,dotsize=\MarkerSize,linecolor=color23.0046]}
}%
}%
} 
\end{pspicture}%

\caption{An example of the utility functions $U_i(r_{i})$ (three sigmoidal-like functions and three logarithmic functions).}
\label{fig:sim:Utilities}
\end{figure}

\textbf{Utility Functions:} The application utility function $U_i(r_i)$ represents the $i^{th}$ user satisfaction percentage with the allocated resources $r_i$. In our model, we assume the utility function $U_i(r_i)$ to be a strictly concave or a sigmoidal-like function. The strictly concave utility function corresponds to the delay-tolerant applications (e.g. FTP, Internet browsing, emails) and sigmoidal-like utility function corresponds to real-time applications (e.g. voice over IP and video streaming). These utility functions have the following properties: (a) $U_i(0) = 0$ and $U_i(r_i)$ is an increasing function of $r_{i}$, and (b) $U_i(r_{i})$ is twice continuously differentiable in $r_{i}$. In our model, we use the normalized sigmoidal-like utility function with $a$ as the rate of increase and $b$ as the inflection point of the function, similar to \cite{DL_PowerAllocation}, and the normalized logarithmic utility function with $k$ as the rate of increase, similar to \cite{UtilityFairness}. Some 
examples of utility functions for different values of $a$, $b$ and $k$ are plotted in Figure \ref{fig:sim:Utilities}. 

\begin{figure}[]
\centering
\includegraphics[width=1.0\linewidth]{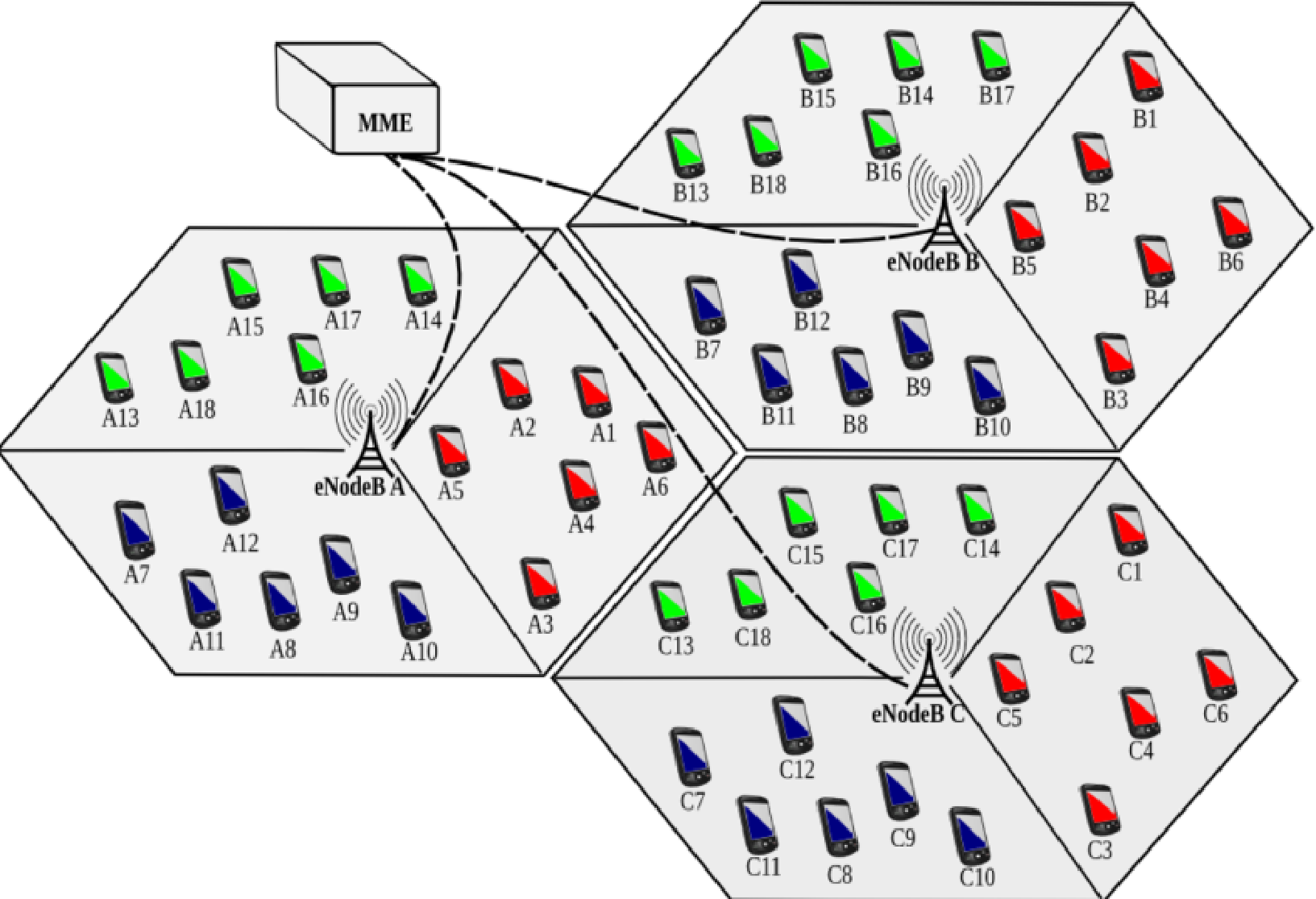}
\caption{Cellular network architecture with 3 cells and 3 sectors per cell. The cells' eNodeBs are connected to Mobility Management Entity (MME).}
\label{fig:System_Model}
\end{figure}

\textbf{System Model:} We consider a cellular network that includes multiple cells where each cell is divided into sectors. The same frequency band is reused for the sector in same direction of different cells. In Figure \ref{fig:System_Model}, we show a diagram of the cellular network architecture consisting of $K$ eNodeBs in $K$ cells, where each cell is divided into $L$ sector (e.g. 3 sectors), and $M$ user equipments (UE)s distributed in these cells. The $i^{th}$ user rate $r_{i}^{l}$ is allocated by the $l^{th}$ sector of eNodeB, where $i = \{1,2, ...,M\}$, $l =\{1,2, ..., L\}$. Each UE has its own utility function $U_i(r_{i})$ that corresponds to the type of application running on it. Our objective is to solve for the optimal rates that eNodeB sectors should allocate to the UEs. The utility functions are given by $U_i(r_{i}^{1}+r_{i}^{2}+ ...+r_{i}^{L})$ where $\sum_{l=1}^{L}r_{i}^{l} = r_i$ and the rate vector is given by $\textbf{r} =\{{r}_1, {r}_2,..., {r}_M\}$. 

\textbf{Resource Allocation Policy:} The context-aware resource allocation policy is achieved by using utility proportional fairness policy, where the objective function is given by the product of utilities $\prod_{i=1}^{M}U_i$. This policy of resource allocation guarantees that the optimal users rates are allocated such that, (a) priority to real-time applications users (i.e. with sigmoidal-like utility functions) over delay-tolerant applications users (i.e. with logarithmic utility functions), (b) no user is dropped (i.e. minimum rate allocation is guaranteed and therefore minimum QoS). 

\textbf{Resource Allocation Optimization Problem:} The basic formulation of the utility proportional fairness resource allocation problem is given by the following optimization problem:
\begin{equation}\label{eqn:opt_prob_fairness}
\begin{aligned}
& \underset{\textbf{r}}{\text{max}} & & \prod_{i=1}^{M}U_i(r_{i}^{1}+r_{i}^{2}+ ...+r_{i}^{L}) \\
& \text{subject to} & & \sum_{l=1}^{L}r_{i}^{l} = r_i ,  \;\; \sum_{i=1}^{M_k}r_{i}^{l} \leq R^{l}, \sum_{l=1}^{L} R^{l} = R, \\
& & &  r_{i}^l \geq 0, \;\;\;\;\;l = 1,2, ...,L, i = 1,2, ...,M.
\end{aligned}
\end{equation}
where $R^{l}$ is the allocated rate by Mobility Management Entity (MME) to the $l^{th}$ sector of all cell, and $M_k$ is the number of users in $k^{th}$ cell. We assume that the cellular network reuses the same frequency band in similar sectors of different cells to avoid interference between cells (i.e. avoid co-channel interference). So we have the assumption that $R^{l}$ is the same for all cells and and $R$ is the sum of the allocated rates to all sectors in a cell. We assume that a UE can't co-exist in two sectors simultaneously.


\section{Context-Aware Global Optimal Solution}\label{sec:Proof}
In this section, we investigate the optimization problem and its dual. Then, we divide the optimization problem into simplified subproblems to be solved in different cellular network entities.

\textbf{Convex Optimization:} In optimization problem (\ref{eqn:opt_prob_fairness}), the solution for the objective function $\prod_{i=1}^{M}U_i$ is equivalent to the solution for the objective function $\sum_{i=1}^{M}\log U_i$. It is shown in \cite{Ahmed_Utility1, Ahmed_Utility4} that utility functions $U_i(.)$ that are logarithmic or sigmoidal-like functions have logarithms $\log U_i(.)$ that are concave functions. Therefore, the optimization problem in (\ref{eqn:opt_prob_fairness}) is convex. It follows that there exists a tractable (i.e can be solved in polynomial time) global optimal solution to the optimization problem in (\ref{eqn:opt_prob_fairness}). 

\textbf{Dual Problem:} The key to a distributed and decentralized optimal solution of the primal problem in (\ref{eqn:opt_prob_fairness}) is to convert it to the dual problem using Lagrangian multiplier methods (more details on how to convert the primal problem to the dual problem can be found in \cite{Ahmed_Utility1, Ahmed_Utility4}). The dual problem of (\ref{eqn:opt_prob_fairness}) can be divided into three simpler subproblems that are solved in UEs, eNodeBs and MME in a bidding process.

\textbf{User Equipment Subproblem:} For the users in the $l^{th}$ sector of the eNodeB, the $i^{th}$ UE optimization problem is given by: 
\begin{equation}\label{eqn:opt_prob_fairness_UE}
r_{i}^{l} = \arg \underset{r_{i}^{l}}\max \Big(\log(U_i(r_{i}^{1}+r_{i}^{2}+ ...+r_{i}^{L}))-\sum_{l=1}^{L}p_lr_{i}^l\Big)
\end{equation}
where $p_l$ is the shadow price (i.e. the price per resource) received from eNodeB $l^{th}$ sector. The shadow price received from eNodeB is computed by solving the eNodeB sector subproblem. In the bidding process, the $i^{th}$ UE maximizes the rate $r_i^{l}$ allocated to it by $l^{th}$ sector of eNodeB by solving the $i^{th}$ UE optimization problem (\ref{eqn:opt_prob_fairness_UE}). The UE bid $w_{i}^l$ corresponding to rate $r_{i}^{l}$ equals $p_l r_{i}^l$.

\textbf{eNodeB Sector Subproblem:} The second subproblem is the eNodeB $l^{th}$ sector optimization problem that minimizes the shadow price $p_l$. The  minimization of $p_l$ is achieved by differentiating the Lagrangian of (\ref{eqn:opt_prob_fairness}) with respect to $p_l$ and setting the inequality constraints in optimization problem (\ref{eqn:opt_prob_fairness}) to equality constraints (i.e setting the slack variable to zero). Finally, we have the $l^{th}$ sector shadow price $p_l = \sfrac{\sum_{i=1}^{M}w_{i}^l}{R^l}$. 

\textbf{MME Subproblem:} For achieving utility proportional fairness policy in context-aware resource allocation optimization problem (\ref{eqn:opt_prob_fairness}), MME performs fair sector total rate $R^{l}$ allocation by setting equal shadow price for all sectors (i.e. $p_l=p$ for all $l$). The aggregated bids from all users in the $l^{th}$ sectors of all eNodeBs is $W^l$ and is equal to $p R^l$. MME computes the sector total rate $R^{l}$ which is proportional to that sector aggregated bid $W^{l}$ to the sum of all sectors aggregated bids and is given by $R^l=\sfrac{W^lR}{\sum_{l=1}^{L}W^l}$ to guarantee utility proportional fairness policy.

\section{Context-Aware Resource Allocation Algorithm}\label{sec:Algorithm}
\begin{figure}[tb]
\centering
\includegraphics[width=1.0\linewidth]{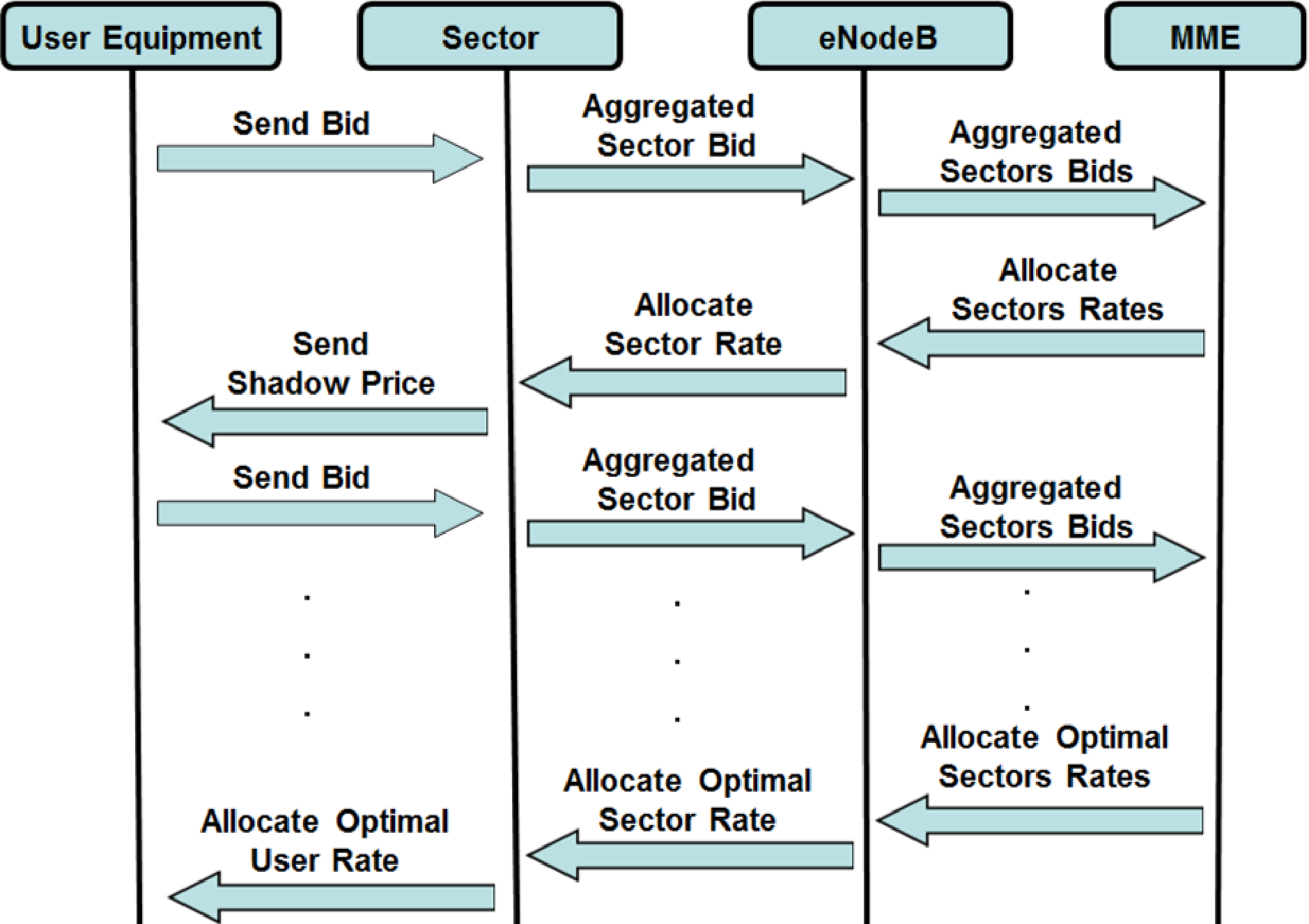}
\caption{The sequence diagram of context-aware resource allocation algorithm that shows the interactions between the different entities of the cellular network.}
\label{fig:sim:signal_flow}
\end{figure}

We translate the cellular entities subproblems into algorithms that run on the corresponding entities. A simplified version of UE, eNodeB and MME algorithms are shown in Algorithm \ref{alg:UE_FK}, \ref{alg:eNodeB_FK}, and \ref{alg:MME_FK}, respectively. The sequence diagram for the context-aware resource allocation algorithm is shown in Figure \ref{fig:sim:signal_flow}. It shows the interactions between UEs, eNodeBs and MME entities.

\begin{algorithm}
\caption{The $i^{th}$ UE in $l^{th}$ sector Algorithm}\label{alg:UE_FK}
\begin{algorithmic}
\STATE {Send initial bid $w_i^{l}(1)$ to $l^{th}$ sector of eNodeB}
\LOOP
	\STATE {Receive shadow price $p_{l}(n)$ from eNodeB}
	\IF {STOP from eNodeB} %
	   	\STATE {STOP}   
	\ELSE
		\STATE {Compute rate $r_{i}^{l}(n)$ using (\ref{eqn:opt_prob_fairness_UE}) and bid $w_i^{l}(n)$}
		\STATE {Send new bid $w_i^{l}(n)$ to $l^{th}$ sector of eNodeB}
	\ENDIF 
\ENDLOOP
\end{algorithmic}
\end{algorithm}

In context-aware resource allocation algorithm, the $i^{th}$ UE starts with an initial bid $w_{i}^l(1)$ which is transmitted to the $l^{th}$ sector of $k^{th}$ eNodeB. The $l^{th}$ sector of $k^{th}$ eNodeB calculates the aggregated sector bid $W_{k}^l(n)$, where $n$ is the time index. The $k^{th}$ eNodeB sends the aggregated bids of all the sectors, $W_{k}^1(n), W_{k}^2(n), ..., W_{k}^L(n)$, to MME. MME adds the aggregated bids of similar sectors of different eNodeBs to evaluate the total aggregated sector bids $W^l(n)$. MME calculates the difference between the total sector aggregated bid $W^l(n)$ and the previously evaluated total sector aggregated bid $W^l(n-1)$ for all sectors and exits if it is less than a pre-specified threshold $\delta$ for all the sectors (i.e. exit criterion). MME evaluates the sector rates $R^l(n)$ for every sector and sends to all eNodeBs. The $l^{th}$ sector of $k^{th}$ eNodeB calculates the shadow price $p_l(n) = \sfrac{\sum_{i=1}^{M}w_{i}^l(n)}{R^l}$ and sends that value to 
all 
the UEs in its coverage area. The $i^{th}$ UE receives the shadow prices $p_{l}$ from corresponding sector to solve the optimization problem in (\ref{eqn:opt_prob_fairness_UE}) and send the new bid  $w_{i}^l(n)$. The bidding process is repeated until $|W^l(n) -W^l(n-1)|$ is less than the threshold $\delta$.

\begin{algorithm}
\caption{The $l^{th}$ sector of eNodeB Algorithm}\label{alg:eNodeB_FK}
\begin{algorithmic}
\LOOP
	\STATE {Receive bids $w_{i}^l(n)$ from UEs}
	\STATE {Calculate aggregated bids $W_k^l(n)$ and send to MME}
	\STATE {Receive sector rate $R^l(n)$ from MME}
	\IF {STOP received from MME} %
		\STATE {STOP and send STOP to all UEs} 
	\ELSE
		\STATE {Calculate $p_l(n)$ and send to all UEs}
	\ENDIF 
\ENDLOOP
\end{algorithmic}
\end{algorithm}

The context-aware resource allocation algorithm is set to avoid the situation of allocating zero rate to any user (i.e. no user is dropped). This is inherited from the utility proportional fairness policy in the optimization problem.  

\begin{algorithm}
\caption{MME Algorithm}\label{alg:MME_FK}
\begin{algorithmic}
\STATE {Send sector rate $R^l(0)$ to $l^{th}$ sector}
\COMMENT{Let $R^l(0) = \frac{R}{L}$}
\LOOP
	\STATE {Receive aggregated bids $W_k^l(n)$ from $l^{th}$ sector}
	\STATE {Calculate total aggregated bids $W^l(n)$}
	\COMMENT{Let $W^l(0) = 0\:\:\forall \:l$}
	\IF {$|W^l(n) - W^l(n-1)|<\delta  \:\:\forall l$} %
		\STATE {STOP and send STOP to all sectors} 
	\ELSE
		\STATE {Calculate $R^l(n)$ and send to $l^{th}$ sector}
	\ENDIF 
\ENDLOOP
\end{algorithmic}
\end{algorithm}
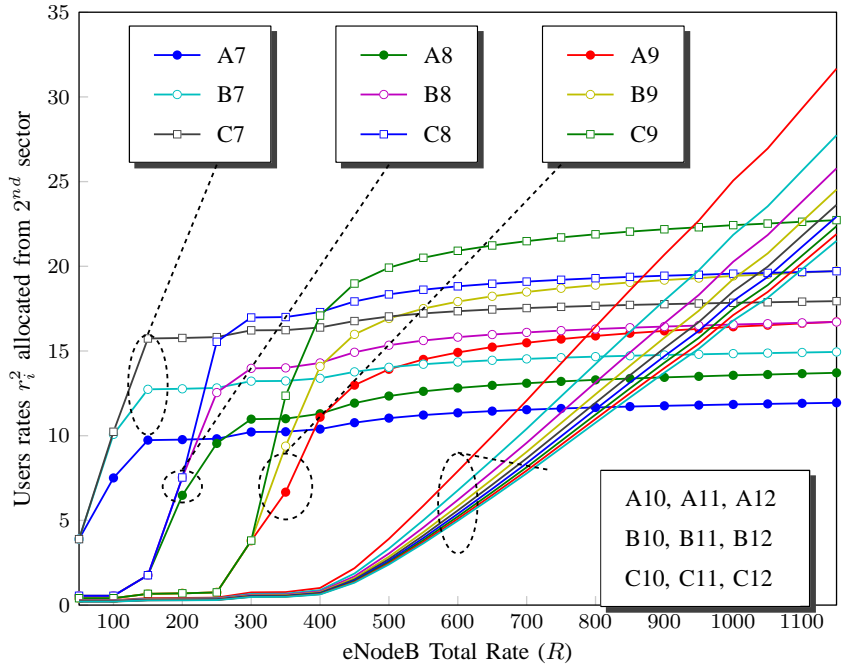
\begin{figure}[tb]
\centering
%
\psset{xunit=0.000909\plotwidth,yunit=0.022535\plotwidth}%
\begin{pspicture}(-61.520737,-3.888889)(1155.069124,35.818713)%


\psline[linewidth=\AxesLineWidth,linecolor=GridColor](100.000000,0.000000)(100.000000,0.532515)
\psline[linewidth=\AxesLineWidth,linecolor=GridColor](200.000000,0.000000)(200.000000,0.532515)
\psline[linewidth=\AxesLineWidth,linecolor=GridColor](300.000000,0.000000)(300.000000,0.532515)
\psline[linewidth=\AxesLineWidth,linecolor=GridColor](400.000000,0.000000)(400.000000,0.532515)
\psline[linewidth=\AxesLineWidth,linecolor=GridColor](500.000000,0.000000)(500.000000,0.532515)
\psline[linewidth=\AxesLineWidth,linecolor=GridColor](600.000000,0.000000)(600.000000,0.532515)
\psline[linewidth=\AxesLineWidth,linecolor=GridColor](700.000000,0.000000)(700.000000,0.532515)
\psline[linewidth=\AxesLineWidth,linecolor=GridColor](800.000000,0.000000)(800.000000,0.532515)
\psline[linewidth=\AxesLineWidth,linecolor=GridColor](900.000000,0.000000)(900.000000,0.532515)
\psline[linewidth=\AxesLineWidth,linecolor=GridColor](1000.000000,0.000000)(1000.000000,0.532515)
\psline[linewidth=\AxesLineWidth,linecolor=GridColor](1100.000000,0.000000)(1100.000000,0.532515)
\psline[linewidth=\AxesLineWidth,linecolor=GridColor](50.000000,0.000000)(63.200000,0.000000)
\psline[linewidth=\AxesLineWidth,linecolor=GridColor](50.000000,5.000000)(63.200000,5.000000)
\psline[linewidth=\AxesLineWidth,linecolor=GridColor](50.000000,10.000000)(63.200000,10.000000)
\psline[linewidth=\AxesLineWidth,linecolor=GridColor](50.000000,15.000000)(63.200000,15.000000)
\psline[linewidth=\AxesLineWidth,linecolor=GridColor](50.000000,20.000000)(63.200000,20.000000)
\psline[linewidth=\AxesLineWidth,linecolor=GridColor](50.000000,25.000000)(63.200000,25.000000)
\psline[linewidth=\AxesLineWidth,linecolor=GridColor](50.000000,30.000000)(63.200000,30.000000)
\psline[linewidth=\AxesLineWidth,linecolor=GridColor](50.000000,35.000000)(63.200000,35.000000)

{ \footnotesize 
\rput[t](100.000000,-0.532515){$100$}
\rput[t](200.000000,-0.532515){$200$}
\rput[t](300.000000,-0.532515){$300$}
\rput[t](400.000000,-0.532515){$400$}
\rput[t](500.000000,-0.532515){$500$}
\rput[t](600.000000,-0.532515){$600$}
\rput[t](700.000000,-0.532515){$700$}
\rput[t](800.000000,-0.532515){$800$}
\rput[t](900.000000,-0.532515){$900$}
\rput[t](1000.000000,-0.532515){$1000$}
\rput[t](1100.000000,-0.532515){$1100$}
\rput[r](36.800000,0.000000){$0$}
\rput[r](36.800000,5.000000){$5$}
\rput[r](36.800000,10.000000){$10$}
\rput[r](36.800000,15.000000){$15$}
\rput[r](36.800000,20.000000){$20$}
\rput[r](36.800000,25.000000){$25$}
\rput[r](36.800000,30.000000){$30$}
\rput[r](36.800000,35.000000){$35$}
} 

\psframe[linewidth=\AxesLineWidth,dimen=middle](50.000000,0.000000)(1150.000000,35.000000)

{ \small 
\rput[b](600.000000,-3.888889){
\begin{tabular}{c}
eNodeB Total Rate ($R$)\\
\end{tabular}
}

\rput[t]{90}(-61.520737,17.500000){
\begin{tabular}{c}
Users rates $r_i^2$ allocated from $2^{nd}$ sector\\
\end{tabular}
}
} 

\newrgbcolor{color548.0023}{0  0  1}
\psline[plotstyle=line,linejoin=1,showpoints=true,dotstyle=*,dotsize=\MarkerSize,linestyle=solid,linewidth=\LineWidth,linecolor=color548.0023]
(50.000000,3.893990)(100.000000,7.506888)(150.000000,9.738176)(200.000000,9.768826)(250.000000,9.819436)
(300.000000,10.219774)(350.000000,10.232818)(400.000000,10.388598)(450.000000,10.768139)(500.000000,11.037309)
(550.000000,11.218498)(600.000000,11.350100)(650.000000,11.451627)(700.000000,11.534533)(750.000000,11.606269)
(800.000000,11.667027)(850.000000,11.719334)(900.000000,11.764722)(950.000000,11.803985)(1000.000000,11.845678)
(1050.000000,11.876244)(1100.000000,11.911513)(1150.000000,11.943659)

\newrgbcolor{color549.0018}{0         0.5           0}
\psline[plotstyle=line,linejoin=1,showpoints=true,dotstyle=*,dotsize=\MarkerSize,linestyle=solid,linewidth=\LineWidth,linecolor=color549.0018]
(50.000000,0.549927)(100.000000,0.549431)(150.000000,1.763434)(200.000000,6.480609)(250.000000,9.545301)
(300.000000,10.977320)(350.000000,11.003536)(400.000000,11.295460)(450.000000,11.923878)(500.000000,12.341977)
(550.000000,12.618510)(600.000000,12.818043)(650.000000,12.971487)(700.000000,13.096557)(750.000000,13.204647)
(800.000000,13.296123)(850.000000,13.374828)(900.000000,13.443093)(950.000000,13.502129)(1000.000000,13.564799)
(1050.000000,13.610735)(1100.000000,13.663729)(1150.000000,13.712022)

\newrgbcolor{color550.0018}{1  0  0}
\psline[plotstyle=line,linejoin=1,showpoints=true,dotstyle=*,dotsize=\MarkerSize,linestyle=solid,linewidth=\LineWidth,linecolor=color550.0018]
(50.000000,0.405774)(100.000000,0.405526)(150.000000,0.664169)(200.000000,0.693017)(250.000000,0.749190)
(300.000000,3.797139)(350.000000,6.667527)(400.000000,11.090820)(450.000000,12.983132)(500.000000,13.920078)
(550.000000,14.503807)(600.000000,14.916238)(650.000000,15.230253)(700.000000,15.484758)(750.000000,15.703916)
(800.000000,15.888924)(850.000000,16.047822)(900.000000,16.185467)(950.000000,16.304384)(1000.000000,16.430521)
(1050.000000,16.522914)(1100.000000,16.629448)(1150.000000,16.726484)

\newrgbcolor{color551.0018}{0        0.75        0.75}
\psline[plotstyle=line,linejoin=1,showpoints=true,dotstyle=Bo,dotsize=\MarkerSize,linestyle=solid,linewidth=\LineWidth,linecolor=color551.0018]
(50.000000,3.893990)(100.000000,10.077854)(150.000000,12.738176)(200.000000,12.768826)(250.000000,12.819436)
(300.000000,13.219774)(350.000000,13.232818)(400.000000,13.388598)(450.000000,13.768139)(500.000000,14.037309)
(550.000000,14.218498)(600.000000,14.350100)(650.000000,14.451627)(700.000000,14.534533)(750.000000,14.606269)
(800.000000,14.667027)(850.000000,14.719334)(900.000000,14.764722)(950.000000,14.803985)(1000.000000,14.845678)
(1050.000000,14.876244)(1100.000000,14.911513)(1150.000000,14.943659)

\newrgbcolor{color552.0018}{0.75           0        0.75}
\psline[plotstyle=line,linejoin=1,showpoints=true,dotstyle=Bo,dotsize=\MarkerSize,linestyle=solid,linewidth=\LineWidth,linecolor=color552.0018]
(50.000000,0.549927)(100.000000,0.549431)(150.000000,1.763434)(200.000000,7.532186)(250.000000,12.545300)
(300.000000,13.977320)(350.000000,14.003536)(400.000000,14.295460)(450.000000,14.923878)(500.000000,15.341977)
(550.000000,15.618510)(600.000000,15.818043)(650.000000,15.971487)(700.000000,16.096557)(750.000000,16.204647)
(800.000000,16.296123)(850.000000,16.374828)(900.000000,16.443093)(950.000000,16.502129)(1000.000000,16.564799)
(1050.000000,16.610735)(1100.000000,16.663729)(1150.000000,16.712022)

\newrgbcolor{color553.0018}{0.75        0.75           0}
\psline[plotstyle=line,linejoin=1,showpoints=true,dotstyle=Bo,dotsize=\MarkerSize,linestyle=solid,linewidth=\LineWidth,linecolor=color553.0018]
(50.000000,0.405775)(100.000000,0.405527)(150.000000,0.664174)(200.000000,0.693022)(250.000000,0.749197)
(300.000000,3.808279)(350.000000,9.382323)(400.000000,14.090749)(450.000000,15.983121)(500.000000,16.920070)
(550.000000,17.503800)(600.000000,17.916231)(650.000000,18.230246)(700.000000,18.484752)(750.000000,18.703910)
(800.000000,18.888918)(850.000000,19.047816)(900.000000,19.185461)(950.000000,19.304378)(1000.000000,19.430515)
(1050.000000,19.522908)(1100.000000,19.629442)(1150.000000,19.726478)

\newrgbcolor{color554.0018}{0.25        0.25        0.25}
\psline[plotstyle=line,linejoin=1,showpoints=true,dotstyle=Bsquare,dotsize=\MarkerSize,linestyle=solid,linewidth=\LineWidth,linecolor=color554.0018]
(50.000000,3.893990)(100.000000,10.231859)(150.000000,15.738176)(200.000000,15.768826)(250.000000,15.819436)
(300.000000,16.219774)(350.000000,16.232818)(400.000000,16.388598)(450.000000,16.768139)(500.000000,17.037309)
(550.000000,17.218498)(600.000000,17.350099)(650.000000,17.451627)(700.000000,17.534533)(750.000000,17.606269)
(800.000000,17.667027)(850.000000,17.719334)(900.000000,17.764722)(950.000000,17.803985)(1000.000000,17.845679)
(1050.000000,17.876244)(1100.000000,17.911514)(1150.000000,17.943660)

\newrgbcolor{color555.0018}{0  0  1}
\psline[plotstyle=line,linejoin=1,showpoints=true,dotstyle=Bsquare,dotsize=\MarkerSize,linestyle=solid,linewidth=\LineWidth,linecolor=color555.0018]
(50.000000,0.549927)(100.000000,0.549431)(150.000000,1.763434)(200.000000,7.532186)(250.000000,15.545300)
(300.000000,16.977320)(350.000000,17.003536)(400.000000,17.295460)(450.000000,17.923878)(500.000000,18.341977)
(550.000000,18.618510)(600.000000,18.818043)(650.000000,18.971487)(700.000000,19.096557)(750.000000,19.204647)
(800.000000,19.296123)(850.000000,19.374828)(900.000000,19.443093)(950.000000,19.502129)(1000.000000,19.564799)
(1050.000000,19.610735)(1100.000000,19.663729)(1150.000000,19.712022)

\newrgbcolor{color556.0018}{0         0.5           0}
\psline[plotstyle=line,linejoin=1,showpoints=true,dotstyle=Bsquare,dotsize=\MarkerSize,linestyle=solid,linewidth=\LineWidth,linecolor=color556.0018]
(50.000000,0.405775)(100.000000,0.405527)(150.000000,0.664174)(200.000000,0.693023)(250.000000,0.749198)
(300.000000,3.808843)(350.000000,12.359918)(400.000000,17.090745)(450.000000,18.983120)(500.000000,19.920070)
(550.000000,20.503800)(600.000000,20.916231)(650.000000,21.230246)(700.000000,21.484752)(750.000000,21.703909)
(800.000000,21.888917)(850.000000,22.047816)(900.000000,22.185460)(950.000000,22.304378)(1000.000000,22.430515)
(1050.000000,22.522908)(1100.000000,22.629442)(1150.000000,22.726477)

\newrgbcolor{color557.0018}{1  0  0}
\psline[plotstyle=line,linejoin=1,linestyle=solid,linewidth=\LineWidth,linecolor=color557.0018]
(50.000000,0.294021)(100.000000,0.293890)(150.000000,0.410625)(200.000000,0.421484)(250.000000,0.441588)
(300.000000,0.749038)(350.000000,0.765486)(400.000000,1.009749)(450.000000,2.177172)(500.000000,3.913196)
(550.000000,5.866330)(600.000000,7.904919)(650.000000,9.973217)(700.000000,12.075674)(750.000000,14.265006)
(800.000000,16.440278)(850.000000,18.587887)(900.000000,20.686774)(950.000000,22.700335)(1000.000000,25.062016)
(1050.000000,26.954131)(1100.000000,29.322461)(1150.000000,31.668800)

\newrgbcolor{color558.0018}{0        0.75        0.75}
\psline[plotstyle=line,linejoin=1,linestyle=solid,linewidth=\LineWidth,linecolor=color558.0018]
(50.000000,0.270745)(100.000000,0.270630)(150.000000,0.372200)(200.000000,0.381572)(250.000000,0.398895)
(300.000000,0.660873)(350.000000,0.674790)(400.000000,0.880896)(450.000000,1.864119)(500.000000,3.337741)
(550.000000,5.011901)(600.000000,6.772840)(650.000000,8.569982)(700.000000,10.405271)(750.000000,12.323696)
(800.000000,14.235903)(850.000000,16.128845)(900.000000,17.983032)(950.000000,19.765269)(1000.000000,21.859426)
(1050.000000,23.539865)(1100.000000,25.646249)(1150.000000,27.736070)

\newrgbcolor{color559.0018}{0.75           0        0.75}
\psline[plotstyle=line,linejoin=1,linestyle=solid,linewidth=\LineWidth,linecolor=color559.0018]
(50.000000,0.254540)(100.000000,0.254434)(150.000000,0.346956)(200.000000,0.355467)(250.000000,0.371192)
(300.000000,0.608283)(350.000000,0.620860)(400.000000,0.807130)(450.000000,1.698550)(500.000000,3.044166)
(550.000000,4.582502)(600.000000,6.207898)(650.000000,7.872223)(700.000000,9.576225)(750.000000,11.361134)
(800.000000,13.143346)(850.000000,14.910147)(900.000000,16.642882)(950.000000,18.310099)(1000.000000,20.271008)
(1050.000000,21.845853)(1100.000000,23.821381)(1150.000000,25.782874)

\newrgbcolor{color560.0018}{0.75        0.75           0}
\psline[plotstyle=line,linejoin=1,linestyle=solid,linewidth=\LineWidth,linecolor=color560.0018]
(50.000000,0.242267)(100.000000,0.242168)(150.000000,0.328474)(200.000000,0.336404)(250.000000,0.351052)
(300.000000,0.571734)(350.000000,0.583441)(400.000000,0.756917)(450.000000,1.590011)(500.000000,2.854712)
(550.000000,4.307024)(600.000000,5.846322)(650.000000,7.426053)(700.000000,9.046244)(750.000000,10.745747)
(800.000000,12.444658)(850.000000,14.130507)(900.000000,15.785198)(950.000000,17.378421)(1000.000000,19.253516)
(1050.000000,20.760292)(1100.000000,22.651393)(1150.000000,24.530016)

\newrgbcolor{color561.0018}{0.25        0.25        0.25}
\psline[plotstyle=line,linejoin=1,linestyle=solid,linewidth=\LineWidth,linecolor=color561.0018]
(50.000000,0.232482)(100.000000,0.232389)(150.000000,0.314069)(200.000000,0.321570)(250.000000,0.335424)
(300.000000,0.544180)(350.000000,0.555260)(400.000000,0.719546)(450.000000,1.511030)(500.000000,2.718058)
(550.000000,4.108908)(600.000000,5.586536)(650.000000,7.105545)(700.000000,8.665470)(750.000000,10.303467)
(800.000000,11.942311)(850.000000,13.569723)(900.000000,15.168022)(950.000000,16.707734)(1000.000000,18.520723)
(1050.000000,19.978202)(1100.000000,21.808121)(1150.000000,23.626651)

\newrgbcolor{color619.0002}{0  0  1}
\psline[plotstyle=line,linejoin=1,linestyle=solid,linewidth=\LineWidth,linecolor=color619.0002]
(50.000000,0.224403)(100.000000,0.224314)(150.000000,0.302368)(200.000000,0.309535)(250.000000,0.322772)
(300.000000,0.522320)(350.000000,0.532918)(400.000000,0.690158)(450.000000,1.449851)(500.000000,2.612791)
(550.000000,3.956545)(600.000000,5.386815)(650.000000,6.859114)(700.000000,8.372614)(750.000000,9.963174)
(800.000000,11.555644)(850.000000,13.137900)(900.000000,14.692591)(950.000000,16.190898)(1000.000000,17.955801)
(1050.000000,19.375091)(1100.000000,21.157590)(1150.000000,22.929523)

\newrgbcolor{color620.0002}{0         0.5           0}
\psline[plotstyle=line,linejoin=1,linestyle=solid,linewidth=\LineWidth,linecolor=color620.0002]
(50.000000,0.217562)(100.000000,0.217476)(150.000000,0.292582)(200.000000,0.299478)(250.000000,0.312216)
(300.000000,0.504355)(350.000000,0.514567)(400.000000,0.666163)(450.000000,1.400439)(500.000000,2.528087)
(550.000000,3.834058)(600.000000,5.226267)(650.000000,6.660970)(700.000000,8.137060)(750.000000,9.689356)
(800.000000,11.244388)(850.000000,12.790163)(900.000000,14.309603)(950.000000,15.774421)(1000.000000,17.500413)
(1050.000000,18.888786)(1100.000000,20.632881)(1150.000000,22.367061)

\newrgbcolor{color621.0002}{1  0  0}
\psline[plotstyle=line,linejoin=1,linestyle=solid,linewidth=\LineWidth,linecolor=color621.0002]
(50.000000,0.211655)(100.000000,0.211572)(150.000000,0.284216)(200.000000,0.290887)(250.000000,0.303210)
(300.000000,0.489206)(350.000000,0.499099)(400.000000,0.646031)(450.000000,1.359315)(500.000000,2.457776)
(550.000000,3.732439)(600.000000,5.093058)(650.000000,6.496517)(700.000000,7.941486)(750.000000,9.461925)
(800.000000,10.985765)(850.000000,12.501128)(900.000000,13.991165)(950.000000,15.428038)(1000.000000,17.121546)
(1050.000000,18.484101)(1100.000000,20.196114)(1150.000000,21.898744)

\newrgbcolor{color622.0002}{0        0.75        0.75}
\psline[plotstyle=line,linejoin=1,linestyle=solid,linewidth=\LineWidth,linecolor=color622.0002]
(50.000000,0.206478)(100.000000,0.206397)(150.000000,0.276941)(200.000000,0.283420)(250.000000,0.295390)
(300.000000,0.476177)(350.000000,0.485800)(400.000000,0.628784)(450.000000,1.324309)(500.000000,2.398037)
(550.000000,3.646124)(600.000000,4.979887)(650.000000,6.356757)(700.000000,7.775218)(750.000000,9.268501)
(800.000000,10.765736)(850.000000,12.255146)(900.000000,13.720080)(950.000000,15.133087)(1000.000000,16.798841)
(1050.000000,18.139329)(1100.000000,19.823915)(1150.000000,21.499566)

{ \small 
\rput(230,30){%
\psshadowbox[framesep=0pt,linewidth=\AxesLineWidth]{\psframebox*{\begin{tabular}{l}
\Rnode{a1}{\hspace*{0.0ex}} \hspace*{0.4cm} \Rnode{a2}{~~A7} \\
\Rnode{a7}{\hspace*{0.0ex}} \hspace*{0.4cm} \Rnode{a8}{~~B7} \\
\Rnode{a13}{\hspace*{0.0ex}} \hspace*{0.4cm} \Rnode{a14}{~~C7} \\
\end{tabular}}
\ncline[linestyle=solid,linewidth=\LineWidth,linecolor=color548.0023]{a1}{a2} \ncput{\psdot[dotstyle=*,dotsize=\MarkerSize,linecolor=color548.0023]}
\ncline[linestyle=solid,linewidth=\LineWidth,linecolor=color551.0018]{a7}{a8} \ncput{\psdot[dotstyle=Bo,dotsize=\MarkerSize,linecolor=color551.0018]}
\ncline[linestyle=solid,linewidth=\LineWidth,linecolor=color554.0018]{a13}{a14} \ncput{\psdot[dotstyle=Bsquare,dotsize=\MarkerSize,linecolor=color554.0018]}
}%
}%
} 
{ \small 
\rput(530,30){%
\psshadowbox[framesep=0pt,linewidth=\AxesLineWidth]{\psframebox*{\begin{tabular}{l}
\Rnode{a3}{\hspace*{0.0ex}} \hspace*{0.4cm} \Rnode{a4}{~~A8} \\
\Rnode{a9}{\hspace*{0.0ex}} \hspace*{0.4cm} \Rnode{a10}{~~B8} \\
\Rnode{a15}{\hspace*{0.0ex}} \hspace*{0.4cm} \Rnode{a16}{~~C8} \\
\end{tabular}}
\ncline[linestyle=solid,linewidth=\LineWidth,linecolor=color549.0018]{a3}{a4} \ncput{\psdot[dotstyle=*,dotsize=\MarkerSize,linecolor=color549.0018]}
\ncline[linestyle=solid,linewidth=\LineWidth,linecolor=color552.0018]{a9}{a10} \ncput{\psdot[dotstyle=Bo,dotsize=\MarkerSize,linecolor=color552.0018]}
\ncline[linestyle=solid,linewidth=\LineWidth,linecolor=color555.0018]{a15}{a16} \ncput{\psdot[dotstyle=Bsquare,dotsize=\MarkerSize,linecolor=color555.0018]}
}%
}%
} 
{ \small 
\rput(830,30){%
\psshadowbox[framesep=0pt,linewidth=\AxesLineWidth]{\psframebox*{\begin{tabular}{l}
\Rnode{a5}{\hspace*{0.0ex}} \hspace*{0.4cm} \Rnode{a6}{~~A9} \\
\Rnode{a11}{\hspace*{0.0ex}} \hspace*{0.4cm} \Rnode{a12}{~~B9} \\
\Rnode{a17}{\hspace*{0.0ex}} \hspace*{0.4cm} \Rnode{a18}{~~C9} \\
\end{tabular}}
\ncline[linestyle=solid,linewidth=\LineWidth,linecolor=color550.0018]{a5}{a6} \ncput{\psdot[dotstyle=*,dotsize=\MarkerSize,linecolor=color550.0018]}
\ncline[linestyle=solid,linewidth=\LineWidth,linecolor=color553.0018]{a11}{a12} \ncput{\psdot[dotstyle=Bo,dotsize=\MarkerSize,linecolor=color553.0018]}
\ncline[linestyle=solid,linewidth=\LineWidth,linecolor=color556.0018]{a17}{a18} \ncput{\psdot[dotstyle=Bsquare,dotsize=\MarkerSize,linecolor=color556.0018]}
}%
}%
} 
{ \small 
\rput[tr](1120,8){%
\psshadowbox[framesep=0pt,linewidth=\AxesLineWidth]{\psframebox*{\begin{tabular}{l}
A10, A11, A12\\
B10, B11, B12\\
C10, C11, C12\\
\end{tabular}}
}%
}%
} 
\psellipse[linewidth=\LineWidth,linestyle=dashed,dash=0.06cm 0.06cm](150,13)(30,3)
\psline[linewidth=\LineWidth,linestyle=dashed,dash=0.06cm 0.06cm](150,16)(250,26)
\psellipse[linewidth=\LineWidth,linestyle=dashed,dash=0.06cm 0.06cm](200,7)(30,1)
\psline[linewidth=\LineWidth,linestyle=dashed,dash=0.06cm 0.06cm](200,8)(500,26)
\psellipse[linewidth=\LineWidth,linestyle=dashed,dash=0.06cm 0.06cm](350,7)(40,2)
\psline[linewidth=\LineWidth,linestyle=dashed,dash=0.06cm 0.06cm](350,9)(750,26)
\psellipse[linewidth=\LineWidth,linestyle=dashed,dash=0.06cm 0.06cm](600,6)(30,3)
\psline[linewidth=\LineWidth,linestyle=dashed,dash=0.06cm 0.06cm](600,9)(730,8)
\end{pspicture}%
\caption{The users rates allocated $r_{i}^{2}$ from $2^{nd}$ sector of all cells.}
\label{fig:sim:rates}
\end{figure}
\begin{figure}
\centering
%
\psset{xunit=0.000909\plotwidth,yunit=0.0013\plotwidth}%
\begin{pspicture}(-79.262673,-66.666667)(1155.069124,614.035088)%


\psline[linewidth=\AxesLineWidth,linecolor=GridColor](100.000000,0.000000)(100.000000,9.128834)
\psline[linewidth=\AxesLineWidth,linecolor=GridColor](200.000000,0.000000)(200.000000,9.128834)
\psline[linewidth=\AxesLineWidth,linecolor=GridColor](300.000000,0.000000)(300.000000,9.128834)
\psline[linewidth=\AxesLineWidth,linecolor=GridColor](400.000000,0.000000)(400.000000,9.128834)
\psline[linewidth=\AxesLineWidth,linecolor=GridColor](500.000000,0.000000)(500.000000,9.128834)
\psline[linewidth=\AxesLineWidth,linecolor=GridColor](600.000000,0.000000)(600.000000,9.128834)
\psline[linewidth=\AxesLineWidth,linecolor=GridColor](700.000000,0.000000)(700.000000,9.128834)
\psline[linewidth=\AxesLineWidth,linecolor=GridColor](800.000000,0.000000)(800.000000,9.128834)
\psline[linewidth=\AxesLineWidth,linecolor=GridColor](900.000000,0.000000)(900.000000,9.128834)
\psline[linewidth=\AxesLineWidth,linecolor=GridColor](1000.000000,0.000000)(1000.000000,9.128834)
\psline[linewidth=\AxesLineWidth,linecolor=GridColor](1100.000000,0.000000)(1100.000000,9.128834)
\psline[linewidth=\AxesLineWidth,linecolor=GridColor](50.000000,0.000000)(63.200000,0.000000)
\psline[linewidth=\AxesLineWidth,linecolor=GridColor](50.000000,100.000000)(63.200000,100.000000)
\psline[linewidth=\AxesLineWidth,linecolor=GridColor](50.000000,200.000000)(63.200000,200.000000)
\psline[linewidth=\AxesLineWidth,linecolor=GridColor](50.000000,300.000000)(63.200000,300.000000)
\psline[linewidth=\AxesLineWidth,linecolor=GridColor](50.000000,400.000000)(63.200000,400.000000)
\psline[linewidth=\AxesLineWidth,linecolor=GridColor](50.000000,500.000000)(63.200000,500.000000)
\psline[linewidth=\AxesLineWidth,linecolor=GridColor](50.000000,600.000000)(63.200000,600.000000)

{ \footnotesize 
\rput[t](100.000000,-9.128834){$100$}
\rput[t](200.000000,-9.128834){$200$}
\rput[t](300.000000,-9.128834){$300$}
\rput[t](400.000000,-9.128834){$400$}
\rput[t](500.000000,-9.128834){$500$}
\rput[t](600.000000,-9.128834){$600$}
\rput[t](700.000000,-9.128834){$700$}
\rput[t](800.000000,-9.128834){$800$}
\rput[t](900.000000,-9.128834){$900$}
\rput[t](1000.000000,-9.128834){$1000$}
\rput[t](1100.000000,-9.128834){$1100$}
\rput[r](36.800000,0.000000){$0$}
\rput[r](36.800000,100.000000){$100$}
\rput[r](36.800000,200.000000){$200$}
\rput[r](36.800000,300.000000){$300$}
\rput[r](36.800000,400.000000){$400$}
\rput[r](36.800000,500.000000){$500$}
\rput[r](36.800000,600.000000){$600$}
} 

\psframe[linewidth=\AxesLineWidth,dimen=middle](50.000000,0.000000)(1150.000000,600.000000)

{ \small 
\rput[b](600.000000,-76.666667){
\begin{tabular}{c}
eNodeB Total Rate ($R$)\\
\end{tabular}
}

\rput[t]{90}(-79.262673,300.000000){
\begin{tabular}{c}
Sector Rate ($R^{l}$) / Shadow Price ($p$)\\
\end{tabular}
}
} 

\newrgbcolor{color960.0012}{0  0  1}
\psline[plotstyle=line,linejoin=1,showpoints=true,dotstyle=*,dotsize=\MarkerSize,linestyle=solid,linewidth=\LineWidth,linecolor=color960.0012]
(50.000000,17.072431)(100.000000,31.916340)(150.000000,44.507390)(200.000000,60.435913)(250.000000,71.596590)
(300.000000,88.769726)(350.000000,99.035037)(400.000000,115.689077)(450.000000,134.231825)(500.000000,153.561875)
(550.000000,173.065906)(600.000000,192.567070)(650.000000,212.008859)(700.000000,231.436410)(750.000000,250.967281)
(800.000000,270.412179)(850.000000,289.761883)(900.000000,308.992907)(950.000000,328.056457)(1000.000000,347.673854)
(1050.000000,366.506657)(1100.000000,386.074986)(1150.000000,405.572643)

\newrgbcolor{color961.0007}{0         0.5           0}
\psline[plotstyle=line,linejoin=1,showpoints=true,dotstyle=Bo,dotsize=\MarkerSize,linestyle=solid,linewidth=\LineWidth,linecolor=color961.0007]
(50.000000,16.693198)(100.000000,32.837764)(150.000000,48.427812)(200.000000,64.945974)(250.000000,81.461175)
(300.000000,98.125345)(350.000000,115.323108)(400.000000,132.045485)(450.000000,148.660055)(500.000000,165.257001)
(550.000000,181.855487)(600.000000,198.459009)(650.000000,215.067086)(700.000000,231.680130)(750.000000,248.300493)
(800.000000,264.923215)(850.000000,281.547256)(900.000000,298.171185)(950.000000,314.792789)(1000.000000,331.434152)
(1050.000000,348.052906)(1100.000000,364.698194)(1150.000000,381.343984)

\newrgbcolor{color962.0007}{1  0  0}
\psline[plotstyle=line,linejoin=1,showpoints=true,dotstyle=Bsquare,dotsize=\MarkerSize,linestyle=solid,linewidth=\LineWidth,linecolor=color962.0007]
(50.000000,16.234371)(100.000000,35.245896)(150.000000,57.064798)(200.000000,74.618112)(250.000000,96.942235)
(300.000000,113.104930)(350.000000,135.641855)(400.000000,152.265439)(450.000000,167.108120)(500.000000,181.181125)
(550.000000,195.078608)(600.000000,208.973920)(650.000000,222.924055)(700.000000,236.883460)(750.000000,250.732226)
(800.000000,264.664606)(850.000000,278.690860)(900.000000,292.835908)(950.000000,307.150754)(1000.000000,320.891994)
(1050.000000,335.440437)(1100.000000,349.226820)(1150.000000,363.083373)

\newrgbcolor{color936.0006}{0  0  0}
\psline[plotstyle=line,linejoin=1,showpoints=true,dotstyle=Bsquare,dotsize=\MarkerSize,linestyle=solid,linewidth=\LineWidth,linecolor=color936.0006]
(50.000000,299.8140)(100.000000,299.9627)(150.000000,206.0574)(200.000000,200.0249)(250.000000,189.6618)
(300.000000,102.2676)(350.000000,99.6464)(400.000000,71.2848)(450.000000,27.2273)(500.000000,12.7854)
(550.000000,7.5592)(600.000000,5.1357)(650.000000,3.8043)(700.000000,2.9749)(750.000000,2.4035)
(800.000000,2.0057)(850.000000,1.7161)(900.000000,1.4987)(950.000000,1.3329)(1000.000000,1.1768)
(1050.000000,1.0741)(1100.000000,.9666)(1150.000000,.8780)

\newrgbcolor{color227.0122}{0  0  1}
\psline[plotstyle=line,linejoin=1,showpoints=true,dotstyle=*,dotsize=\MarkerSize,linestyle=dashed,linewidth=\LineWidth,linecolor=color227.0122]
(50.000000,15.384789)(100.000000,29.854709)(150.000000,42.148176)(200.000000,57.828009)(250.000000,69.022272)
(300.000000,84.478600)(350.000000,94.441266)(400.000000,109.477899)(450.000000,119.049850)(500.000000,124.617524)
(550.000000,128.253971)(600.000000,130.960468)(650.000000,133.279716)(700.000000,135.196834)(750.000000,137.552590)
(800.000000,138.990312)(850.000000,140.717694)(900.000000,143.000761)(950.000000,142.441802)(1000.000000,145.470186)
(1050.000000,144.813854)(1100.000000,144.318554)(1150.000000,148.459248)

\newrgbcolor{color228.0112}{0         0.5           0}
\psline[plotstyle=line,linejoin=1,showpoints=true,dotstyle=Bo,dotsize=\MarkerSize,linestyle=dashed,linewidth=\LineWidth,linecolor=color228.0112]
(50.000000,17.536764)(100.000000,33.533606)(150.000000,49.611960)(200.000000,66.126178)(250.000000,82.762586)
(300.000000,100.257820)(350.000000,117.508857)(400.000000,135.253650)(450.000000,156.852903)(500.000000,180.984629)
(550.000000,206.245684)(600.000000,231.991999)(650.000000,257.882837)(700.000000,283.972375)(750.000000,309.662264)
(800.000000,335.970858)(850.000000,362.019600)(900.000000,387.632178)(950.000000,415.243658)(1000.000000,440.270251)
(1050.000000,467.878099)(1100.000000,495.334407)(1150.000000,519.536155)

\newrgbcolor{color229.0112}{1  0  0}
\psline[plotstyle=line,linejoin=1,showpoints=true,dotstyle=Bsquare,dotsize=\MarkerSize,linestyle=dashed,linewidth=\LineWidth,linecolor=color229.0112]
(50.000000,17.078447)(100.000000,36.611685)(150.000000,58.239863)(200.000000,76.045813)(250.000000,98.215142)
(300.000000,115.263580)(350.000000,138.049877)(400.000000,155.268451)(450.000000,174.097246)(500.000000,194.397847)
(550.000000,215.500345)(600.000000,237.047533)(650.000000,258.837447)(700.000000,280.830791)(750.000000,302.785146)
(800.000000,325.038830)(850.000000,347.262706)(900.000000,369.367061)(950.000000,392.314540)(1000.000000,414.259563)
(1050.000000,437.308047)(1100.000000,460.347038)(1150.000000,482.004597)

{ \small 
\rput[tl](74.932796,588.885208){%
\psshadowbox[framesep=0pt,linewidth=\AxesLineWidth]{\psframebox*{\begin{tabular}{l}
Balanced traffic\\
\Rnode{a1}{\hspace*{0.0ex}} \hspace*{0.4cm} \Rnode{a2}{~~$R^1$}  \\
\Rnode{a3}{\hspace*{0.0ex}} \hspace*{0.4cm} \Rnode{a4}{~~$R^2$}  \\
\Rnode{a5}{\hspace*{0.0ex}} \hspace*{0.4cm} \Rnode{a6}{~~$R^3$}  \\
\end{tabular}}
\ncline[linestyle=solid,linewidth=\LineWidth,linecolor=color227.0122]{a1}{a2} \ncput{\psdot[dotstyle=*,dotsize=\MarkerSize,linecolor=color227.0122]}
\ncline[linestyle=solid,linewidth=\LineWidth,linecolor=color228.0112]{a3}{a4} \ncput{\psdot[dotstyle=Bo,dotsize=\MarkerSize,linecolor=color228.0112]}
\ncline[linestyle=solid,linewidth=\LineWidth,linecolor=color229.0112]{a5}{a6} \ncput{\psdot[dotstyle=Bsquare,dotsize=\MarkerSize,linecolor=color229.0112]}

}%
}%
} 

{ \small 
\rput[tl](474.932796,588.885208){%
\psshadowbox[framesep=0pt,linewidth=\AxesLineWidth]{\psframebox*{\begin{tabular}{l}
Unbalanced traffic\\
\Rnode{a7}{\hspace*{0.0ex}} \hspace*{0.4cm} \Rnode{a8}{~~$R^1$}\\
\Rnode{a9}{\hspace*{0.0ex}} \hspace*{0.4cm} \Rnode{a10}{~~$R^2$}\\
\Rnode{a11}{\hspace*{0.0ex}} \hspace*{0.4cm} \Rnode{a12}{~~$R^3$}\\
\end{tabular}}
\ncline[linestyle=dashed,linewidth=\LineWidth,linecolor=color227.0122]{a7}{a8} \ncput{\psdot[dotstyle=*,dotsize=\MarkerSize,linecolor=color227.0122]}
\ncline[linestyle=dashed,linewidth=\LineWidth,linecolor=color228.0112]{a9}{a10} \ncput{\psdot[dotstyle=Bo,dotsize=\MarkerSize,linecolor=color228.0112]}
\ncline[linestyle=dashed,linewidth=\LineWidth,linecolor=color229.0112]{a11}{a12} \ncput{\psdot[dotstyle=Bsquare,dotsize=\MarkerSize,linecolor=color229.0112]}
}%
}%
} 

{ \small 
\rput[tl](574.932796,108.885208){%
\psshadowbox[framesep=0pt,linewidth=\AxesLineWidth]{\psframebox*{\begin{tabular}{l}
\Rnode{a13}{\hspace*{0.0ex}} \hspace*{0.4cm} \Rnode{a14}{~~ Shadow Price $p$ }  \\
\Rnode{a1}{\hspace*{0.0ex}} \hspace*{0.4cm} \Rnode{a2}{~~ (scaled by 100)}  \\
\end{tabular}}
\ncline[linestyle=solid,linewidth=\LineWidth,linecolor=color936.0006]{a13}{a14} \ncput{\psdot[dotstyle=Bsquare,dotsize=\MarkerSize,linecolor=color936.0006]}

}%
}%
} 

\end{pspicture}%
\caption{(a) Balanced users traffic in all sectors \{solid lines\}, (b) Unbalanced users traffic in sectors $R^l$ (when users A4, A5, A6, B4, B5, B6, C4, C5 and C6, in Figure \ref{fig:System_Model}, exit the cellular network) \{dashed lines\}, (c) Pricing scaled by 100 for the balanced traffic case \{black line\}.}
\label{fig:sim:shadow_price}
\end{figure}
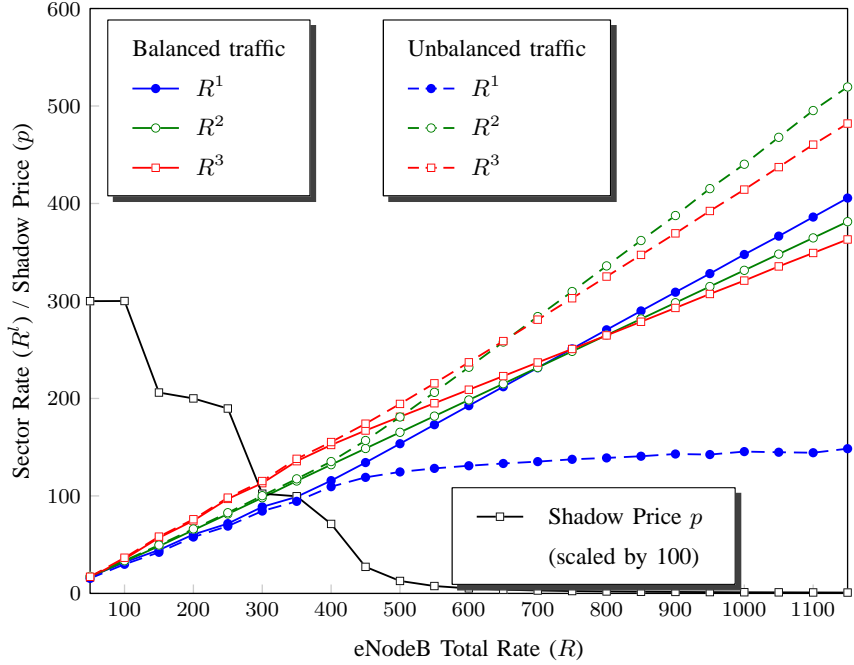
\section{Experimental Results}\label{sec:sim}

In this section, we provide the performance of the context-aware resource allocation algorithm. Algorithm (\ref{alg:UE_FK}), (\ref{alg:eNodeB_FK}) and (\ref{alg:MME_FK}) were applied to various logarithmic and sigmoidal-like utility functions with different parameters in MATLAB. The simulation results showed convergence to the global optimal rates with the desired policy (a) priority to real-time users, and (b) no user is dropped. Figure \ref{fig:System_Model} illustrates the users deployment and indexes in the cellular network sectors used in the simulations. Table \ref{tab:utility} presents the users utilities used in the simulation. In the following simulations, we set $\delta =10^{-3}$ and the eNodeB rate $R$ takes values between 50 and 1150 with step of 5.

\begin {table}[t]
\caption {Users Utilities} 
\label{tab:utility} 
\begin{center}
\begin{tabular}{| l | l || l | l |}
  \hline
  \multicolumn{4}{|c|}{Sector 1 eNodeB A} \\  \hline
  A1 & Sig $a=3,\:\: b=10.0$ & A4 & Log $k=1.1$  \\ \hline
  A2 & Sig $a=3,\:\: b=10.3$ & A5 & Log $k=1.2$ \\ \hline
  A3 & Sig $a=1,\:\: b=10.6$ & A6 & Log $k=1.3$ \\ \hline
  \multicolumn{4}{|c|}{Sector 2 eNodeB A} \\  \hline
  A7 & Sig $a=3,\:\: b=10$ & A10 & Log $k=1$  \\ \hline
  A8 & Sig $a=3,\:\: b=11$ & A11 & Log $k=2$ \\ \hline
  A9 & Sig $a=1,\:\: b=12$ & A12 & Log $k=3$ \\ \hline  
  \multicolumn{4}{|c|}{Sector 3 eNodeB A} \\  \hline
  A13 & Sig $a=3,\:\: b=15.1$ & A16 & Log $k=10$  \\ \hline
  A14 & Sig $a=3,\:\: b=15.3$ & A17 & Log $k=11$ \\ \hline
  A15 & Sig $a=3,\:\: b=15.5$ & A18 & Log $k=12$ \\ \hline
  
  \multicolumn{4}{|c|}{Sector 1 eNodeB B} \\  \hline
  B1 & Sig $a=3,\:\: b=10.9$ & B4 & Log $k=1.4$  \\ \hline
  B2 & Sig $a=3,\:\: b=11.2$ & B5 & Log $k=1.5$ \\ \hline
  B3 & Sig $a=1,\:\: b=11.5$ & B6 & Log $k=1.6$ \\ \hline
  \multicolumn{4}{|c|}{Sector 2 eNodeB B} \\  \hline
  B7 & Sig $a=3,\:\: b=13$ & B10 & Log $k=4$  \\ \hline
  B8 & Sig $a=3,\:\: b=14$ & B11 & Log $k=5$ \\ \hline
  B9 & Sig $a=1,\:\: b=15$ & B12 & Log $k=6$ \\ \hline  
  \multicolumn{4}{|c|}{Sector 3 eNodeB B} \\  \hline
  B13 & Sig $a=3,\:\: b=15.7$ & B16 & Log $k=13$  \\ \hline
  B14 & Sig $a=3,\:\: b=15.9$ & B17 & Log $k=14$ \\ \hline
  B15 & Sig $a=3,\:\: b=17.3$ & B18 & Log $k=15$ \\ \hline
  
  \multicolumn{4}{|c|}{Sector 1 eNodeB C} \\  \hline
  C1 & Sig $a=3,\:\: b=11.8$ & C4 & Log $k=1.7$  \\ \hline
  C2 & Sig $a=3,\:\: b=12.1$ & C5 & Log $k=1.8$ \\ \hline
  C3 & Sig $a=1,\:\: b=12.4$ & C6 & Log $k=1.9$ \\ \hline
  \multicolumn{4}{|c|}{Sector 2 eNodeB C} \\  \hline
  C7 & Sig $a=3,\:\: b=16$ & C10 & Log $k=7$  \\ \hline
  C8 & Sig $a=3,\:\: b=17$ & C11 & Log $k=8$ \\ \hline
  C9 & Sig $a=1,\:\: b=18$ & C12 & Log $k=9$ \\ \hline  
  \multicolumn{4}{|c|}{Sector 3 eNodeB C} \\  \hline
  C13 & Sig $a=3,\:\: b=17.5$ & C16 & Log $k=16$  \\ \hline
  C14 & Sig $a=3,\:\: b=17.7$ & C17 & Log $k=17$ \\ \hline
  C15 & Sig $a=3,\:\: b=17.9$ & C18 & Log $k=18$ \\ \hline  
\end{tabular}
\end{center}
\end {table}

\textbf{Allocated Rates:} In Figure \ref{fig:sim:rates}, we show the optimal rates of users in the $2^{nd}$ sector versus eNodeB rate $R$. The optimal resource allocation is content-aware. The users with real-time application (i.e. sigmoidal-like utilities) are allocated resources first. In real-time applications allocation, the user with the steepest utility function (largest $a$) is allocated first as shown in Figure \ref{fig:sim:rates}.

\textbf{Pricing:} In Figure \ref{fig:sim:shadow_price}, the shadow price $p$ (scaled by 100 for visibility) is plotted versus eNodeB rate $R$. The price per resource increases as the available resources for allocation in sectors are more scarce, small values of $R$, as the number of users is fixed. Similarly, if the available resources are fixed and the number of users or the traffic increases, the price per resource increases. Therefore, we have a traffic-dependent pricing. As the cellular traffic is dependent on the time of using the service (e.g. peak or off-peak traffic hours) then it is time-aware pricing which is intrinsic in the optimization problem solution. Taking advantage of the time-aware pricing, WSPs can flatten the traffic specially during peak hours by setting time-aware resource price. This will incentivize users to use the cellular network during off-peak traffic hours.

\textbf{Balanced Traffic in Sectors:} In this case, users applications rate requirements are very close or the utility parameters are approximately equal. As a result, the  resources allocated by MME to different sectors in the cellular network are approximately equal. In Figure \ref{fig:sim:shadow_price}, we plot the allocated sector rates $R^{l}$ by MME versus the total eNodeB rate $R$.

\textbf{Unbalanced Traffic in Sectors:} In this case, the users in different sectors have different rate requirements. We assume in this scenario that users A4, A5, A6, B4, B5, B6, C4, C5 and C6, shown in Figure \ref{fig:System_Model} with utility parameters shown in Table \ref{tab:utility}, exit the cellular network. Therefore, the  resources allocated by MME to different sectors in the cellular network are not equal. In Figure \ref{fig:sim:shadow_price}, the allocated sector rates $R^{l}$ by MME versus the total eNodeB rate $R$ are shown.


\section{Future Research Directions}\label{sec:sim}

The research directions for further improvement of the context-aware optimal resource allocation architecture are:

\textbf{Time-Aware Pricing:} The pricing results, presented in simulation section, show that solving for optimal rates provides the corresponding users the optimal price of resource (Figure \ref{fig:sim:shadow_price}) which is a time-aware price. An addition to our architecture that take advantage of the time-aware pricing is a pricing model that is available for users ahead of time \cite{pricing_survey}, e.g. day-ahead pricing. The day-ahead pricing is dependent on the history of data-usage in previous days and the expected data-usage in the next day. The user will have the option to restrict some of his applications, especially delay-tolerant ones, from running in certain time periods, especially during congested traffic hours. The advantages on the mobile user side are (a) lower resource price and (b) better QoS, and on the WSP side are (c) lower equipment deployment costs, (d) lower network operation expenses, and (d) providing a more competitive service in the wireless market.  

\textbf{Location-Aware Carrier Aggregation:} In the presence of two cellular networks (e.g. two different LTE WSPs) or two heterogeneous wireless networks (e.g. WiFi and LTE) in the same cell or sector, then carrier aggregation can be performed. In this case, the UE chooses the WSP that provides the lowest price for resources in the bidding process (more details are in \cite{Ahmed_Utility4}). As the user approaches a WiFi AP, he switches from cellular network to minimize the price of resources (with the assumption that WiFi resource price is lower in this scenario) and at the same time the cellular network benefits from decreasing the traffic load especially during congested traffic hours. Hence, WSP can provide better QoS for its subscribers. So, it is a win-win situation.

\textbf{Content-Aware Discrete Resource Allocation:} An extension of this work is to include discrete resource allocation rather than the continuous one. In this case, the optimization problem is reformulated into a discrete optimization problem. That can be solved optimally using Lagrangian relaxation and branch and bound methods. This architecture modification provides better mapping to the current cellular network standards.


\section{Conclusion}\label{sec:conclude}
Context-aware resource allocation is a simple, optimal and efficient method for allocating resources in broadband wireless cellular networks. It provides the foundation for constructing a complete cellular network with all the features in the current standards e.g. carrier aggregation, frequency reuse, GBR, etc. The mathematical formulation of context-aware resource allocation optimization problem with frequency reuse is presented. The new architecture includes content-aware resource allocation and time-aware pricing and can be easily extended to include location-aware carrier aggregation. The description of the algorithms implemented in different cellular network entities and their sequence diagram are presented. The simulation results demonstrate the content-aware resource resource allocation and time-aware pricing. Hence, the cellular spectrum utilization is more efficient and the QoS to mobile users is improved.


\bibliographystyle{ieeetr}
\bibliography{pubs}
\end{document}